\documentclass[esd, manuscript]{copernicus}

\begin{document}
\begin{nolinenumbers}
\title{Evaluation of preCICE (version 3.3.0) in an Earth System Model Regridding Benchmark}

\Author[1][alex.hocks@tum.de, 0009-0002-0666-5233]{Alex}{Hocks}
\Author[2][benjamin.uekermann@ipvs.uni-stuttgart.de, 0000-0002-1314-9969]{Benjamin}{Uekermann}

\affil[1]{Technical University of Munich, School of Computation, Information and Technology, Chair of Scientific Computing, Arcisstr. 21, 80333 Munich}
\affil[2]{University of Stuttgart, Institute for Parallel and Distributed Systems, Universitätsstr. 32, 70569 Stuttgart}

\runningtitle{Evaluation of preCICE (version 3.3.0) in an Earth System Modeling Regridding Benchmark}
\runningauthor{Hocks et al.}

\received{}
\pubdiscuss{}
\revised{}
\accepted{}
\published{}

\firstpage{1}

\maketitle

\begin{abstract}
In Earth System Modeling (ESM), meshes of different models usually do not match, requiring data mapping algorithms implemented in coupling software. \cite{paper} recently introduced a benchmark to evaluate such algorithms and compared implementations in four specialized ESM couplers. In this paper, we assess preCICE, a general-purpose coupling library not limited to ESM, using this benchmark and compare our results to the original study. The generality of preCICE with its larger community offers potential benefits to ESM applications, but the software naturally lacks ESM-specific solutions. We describe necessary pre- and postprocessing steps to make the benchmark tangible for preCICE. Overall, preCICE achieves comparable results; using its radial basis function mapping yields significantly lower errors.
\end{abstract}

\introduction
Earth System Modeling (ESM) simulates interactions among components of the Earth system to understand and predict climate and environmental dynamics. Individual components, such as the atmosphere (ATM) and ocean (SEA), typically use different meshes, requiring data mapping (or regridding) algorithms to exchange information. In partitioned setups using specialized solvers for each component, these algorithms are typically provided by dedicated couplers.
\cite{paper} introduced an ESM benchmark to evaluate and compare such algorithms and couplers, testing four specialized ESM tools -- SCRIP \citep{scrip}, YAC \citep{yac}, ESMF \citep{esmf}, and XIOS \citep{xios} -- on various ATM–SEA mesh pairs using representative test functions.

In this paper, we assess preCICE \citep{preCICEv2}, a general-purpose coupling library, already used in ESM \citep{Abele2025}, but limited to not this domain, using the same benchmark and compare our results to the original study. The generality of preCICE with its large existing user and developer community offers potential benefits, including:
\begin{itemize}
    \item cross-domain knowledge transfer,
    \item flexibility of the coupling approach,
    \item ready-to-use adapters for widely used simulation software, including such also used in ESM, as OpenFOAM \citep{openfoam-adapter} and FEniCS \citep{fenics-adapter},
    \item extensive user documentation and focus on usability,
    \item sustainability of the software development, and 
    \item advanced numerical and HPC methods.
\end{itemize}

However, this generality also means the absence of ESM-specific optimizations in terms of accuracy and efficiency. For example, preCICE operates solely on Cartesian coordinates, which may require additional pre- and postprocessing for geodesic ESM meshes. The objective of this paper is to study the influence of this lack of ESM-tailored algorithms. We introduce the benchmark problem and preCICE in Section \ref{sec:background} and describe the necessary pre- and postprocessing in Section \ref{sec:implementation}. Afterwards, we discuss the benchmark results in Section \ref{sec:results}.
This paper summarizes and significantly extends the student project report of \cite{idp-report}.

\section{Background} \label{sec:background}

We introduce the background of our study in four sections:
the mathematical formulation of the mapping problem in Section \ref{ssec:mapping},
preCICE in Section \ref{ssec:precice},
the benchmark in Section \ref{ssec:benchmark},
and VTK as an important dependency of the pre- and processing in Section \ref{ssec:vtk}.

\subsection{Mapping Problem}
\label{ssec:mapping}

The source mesh (superscript S) and the target mesh (superscript T) are given by their points, 
\begin{equation}
     \qquad 
     \textbf{x}^S_i \in \mathbb{R}^d, \, i\in \{1,\hdots,N^S\} \;\; \mathrm{and} \;\;
     \textbf{x}^T_j \in \mathbb{R}^d, \, j\in \{1,\hdots,N^T\} \,.
\end{equation}
$d \in \{2,3\}$ is the spatial dimension. The mapping problem consists of transfering scalar or vectorial data from the source to the target mesh. For the benchmark, data stems from a synthetic scalar test function $f:\mathbb{R}^d \rightarrow \mathbb{R}$.
We define $\Psi^S = \left(\Psi^S_1, \hdots, \Psi^S_{N^S} \right)^t \in \mathbb{R}^{N^S}$ as the test function evaluated on the source mesh.
We collect all mapped (regridded) values in $R\Psi^S= \left(R\Psi^S_1, \hdots, R\Psi^S_{N^T} \right)^t \in \mathbb{R}^{N^T}$, following the notation of \cite{paper}.
As the mappings considered in the benchmark can be expressed linearly, we can write the data mapping in matrix notation, 
\begin{equation}
    \qquad M \, \Psi^S = R\Psi^S \,
\end{equation}
with the mapping matrix $M\in \mathbb{R}^{N^T \times N^S}$.
Mappings in preCICE can be classified as consistent or conservative. Consistent mappings are used for intensive quantities, such as velocity or temperature. For such quantities, constant data should remain constant, which means that all rows of $M$ sum up to 1. Conservative mappings are for extensive quantities, such as force. Here, the sum of the data values should remain constant, which means that all columns of $M$ sum up to 1. Thus, consistent and conservative mappings can be constructed from each other by transposing the mapping matrix. In ESM, the term conservative mapping is often used differently. Here, still intensive quantities are considered, but whose integral values should be conserved.  To this end, meshes need to be enriched by connectivity. In \cite{paper} a mapping is considered conservative if the integral of the mapped data over a target cell is equal to the integral of the source data over the regions of the source cells that overlap with the target cell. 

To evaluate mappings, we further consider the test function evaluated on the target mesh $\Psi^T$ and integrals approximations $I^S$, $I^T$ on the source and target mesh, respectively, assuming constant data within each cell. 
As accuracy measures we then consider the mean, max, and rms of the (relative) misfit 
\(|R\Psi^S - \Psi^T| / |\Psi^T|\;.\) 
Furthermore, we use 
\begin{equation}
    \qquad
    L_\mathrm{min} :=(\mathrm{min}\Psi^T-\mathrm{min}R\Psi^S)/\mathrm{max}|\Psi^T| \;\; \text{and} \;\;
    L_\mathrm{max} :=(\mathrm{max}R\Psi^S-\mathrm{max}R\Psi^T)/\mathrm{max}|\Psi^T|\; ,
\end{equation}
to evaluate the conservation of minumum and maximum values. Finally, source and target global conservation is defined as 
\begin{equation}
    \qquad
    |I^T(R\Psi^S)-I^S(\Psi^S)|/I^S(\Psi^S)\;\; \text{and} \;\;
    |I^T(R\Psi^S)-I^T(\Psi^T)|/I^T(\Psi^T)\;.
\end{equation}

\subsection{preCICE Ecosystem}
\label{ssec:precice}

Beyond the coupling library, preCICE is a full software ecosystem, comprising adapters to simulation codes, bindings to various programming languages, and a multitude of tools. In this section, we introduce the relevant parts for our study: the core library and the ASTE tool. 

\subsubsection{Library}

The overall vision of preCICE is to enable minimally invasive integration of coupling into existing simulation codes. To this end, preCICE follows a black-box coupling principle: coupling is achieved without information on the underlying physics or the numerics of the simulation codes.
Consistent with preCICE nomenclature, we call such simulation codes \textit{participants} of a coupled simulation.
Each participants defines one or multiples coupling meshes as clouds of vertices in a Cartesian coordinate system without, at first, any connectivity information. The mesh locations can refer to nodes, cell centers, or any other required locations.

The preCICE library implements three fundamental features:
\begin{enumerate}
    \item coupling schemes and accelerations,
    \item communications, and
    \item data mappings.
\end{enumerate}

Coupling schemes determine the scheduling, in which participants contribute towards each time step. Participants can run concurrently (parallel coupling) or sequentially (serial coupling). One can also define the direction of data-flow, i.e., whether participants send data to each other or only uni-directionally. Finally, a choice between explicit and implicit coupling has to be made. While explicit coupling computes each time window only once, implicit coupling sub-iterates. Mathematically, this corresponds to an iterative solution procedure of a fixed-point equation up to a defined threshold. The library offers various acceleration methods for this iterative procedure including Aitken under-relaxation and quasi-Newton methods. If participants use non-matching time step sizes, coupling schemes can be combined with time interpolation \citep{time-interp}.

Inter-participant communication uses a peer-to-peer layout, avoiding any server instances. Based on the different distributed meshes of coupled participants, preCICE computes individual communiation channels during the initialization, a process called repartitioning in preCICE. The actual communication is then parallel and asynchronous using either MPI or TCP/IP as backends.

In terms of data mappings, preCICE offers mainly three different mapping methods: nearest-neighbour (NN), neareast-projection (NP), and a family of radial basis function (RBF) methods. 
In their consistent variants, NN does a nearest-neighbor search from the target to the source mesh during initialization and then copies data values from the source to the target in each time step or iteration.
NP projects target mesh nodes onto the nearest mesh element (e.g., surface triangle) of the source mesh during initialization. This means, that to use this method, mesh connectivity information needs to be added to the source mesh. In each time step or iteration, values on the source mesh are linearly interpolated on the projection point and then copied to the target mesh.
Finally, as RBF method, we focus on the partition-of-unity variant implemented in preCICE \citep{Schneider2025}. This method partitions the distributed source meshes into clusters, for each of which an RBF interpolant is computed by solving a small linear system. For each target mesh node, associated clusters are computed and interpolants are glued together if necessary.
All three mapping methods can be configured to be consistent or conservative. As a special case, preCICE offers \textit{scaled-consistent} as a third experimental option. Here, the results of a consistent mapping are globally scaled in a post-processing step to conserve integral values. To compute the scaling factor, connectivity information of both meshes is required.

\subsubsection{ASTE}

The artificial solver testing environment (ASTE) \citep{aste} can be used to evaluate mappings on complex geometries. It is a lightweight wrapper around the preCICE API and emulates coupled participants. Besides this core functionality, a number of helper scripts are included to partition or join meshes or to evaluate test functions on meshes.

\subsection{Benchmark Setup}
\label{ssec:benchmark}

To introduce the benchmark, we closely follow the notation and naming schemes of the original paper by \cite{paper}. The input data is technically made available in said paper. However, due to uncertainties regarding file and data naming, we obtained it directly from the authors. The original meshes are provided in NetCDF (NC) files, a self-contained binary data format including meta data \citep{netcdf}.
Besides the results, we also provide the input data as part of a data publication \cite{data} to further improve the re-usability of the benchmark.

\subsubsection{Meshes}

As Earth is not a perfect sphere, we use the WGS84 model \citep{wgs}, resulting in an ellipsoid. Our goal is to study data mapping between ATM and SEA models. We, thus, only consider the boundary of the ellipsoid. ATM models are likely to cover the complete shell around the ellipsoid, while SEA models may only contain a subsection of it, the regions of oceans, masking out the rest. The resulting meshes reflect different discretization approaches of these models. 
All meshes are discretized into cells of not necessarily uniform geometry and number of points. Meshes are provided in two forms: a) node-based including connectivity and b) cell-based, only containing cell centers. Cell areas are computed from the node-based form.

The SEA meshes are:
\begin{itemize}
    \item torc: ocean model NEMO\footnote{\label{fn:nemo}\url{https://www.nemo-ocean.eu/}} (Nucleus for European Modelling of the Ocean), low resolution quadrilateral
    \item nogt: ocean model NEMO, higher resolution quadrilateral
\end{itemize}
The ATM meshes are:
\begin{itemize}
    \item bggd: atmosphere model LMDz\footnote{\label{fn:lmdz}\url{https://forge.ipsl.jussieu.fr/igcmg_doc/wiki/Doc/Models/LMDZ}} (Laboratoire de Météorologie Dynamique zoom), only quadrilaterals
    \item sse7: atmosphere model ARPEGE\footnote{\label{fn:arpege}\url{https://www.umr-cnrm.fr/spip.php?article124}} (Action de Recherche Petite Echelle Grande Echelle), mostly quadrilaterals, but has certain regions in which cells contains up to 7 points
    \item icoh: atmosphere model Dynamico\footnote{\label{fn:dynamico}\url{https://www.lmd.polytechnique.fr/~dubos/DYNAMICO/}}, hexa- and pentagons with very high resolution
    \item icos: atmosphere model Dynamico, hexa- and pentagons with low resolution
\end{itemize}

We refer to \cite{paper} for more details and the origins of the meshes.
ESM models typically use 2D geodesic representations using lat/lon coordinates, which is also the case for the meshes of the original benchmark. To use preCICE for coupling, we need to convert these into 3D Cartesian representations using x/y/z coordinates.

\subsubsection{Test Functions}
To provide test data for mapping, four test functions are used:
\begin{itemize}
    \item sinusoid: slowly varying standard sinusoid over the globe,
    \item harmonic: more rapidly varying function with 16 maxima and 16 minima in northern and southern bands,
    \item vortex: slowly varying function with two added vortices, over the Atlantic and Indonesia, and
    \item gulfstream: slowly varying standard sinusoid with a mimicked Gulf Stream
\end{itemize}

In the benchmark, the functions are evaluated using geodesic coordinates, see Figure \ref{fig:eval-nogt} for a visualization on the nogt mesh. For the exact definitions of the functions, we refer to the original work \citep{paper}.

\begin{figure*}[hbt!]
    \centering
    \includegraphics[width=0.85\linewidth]{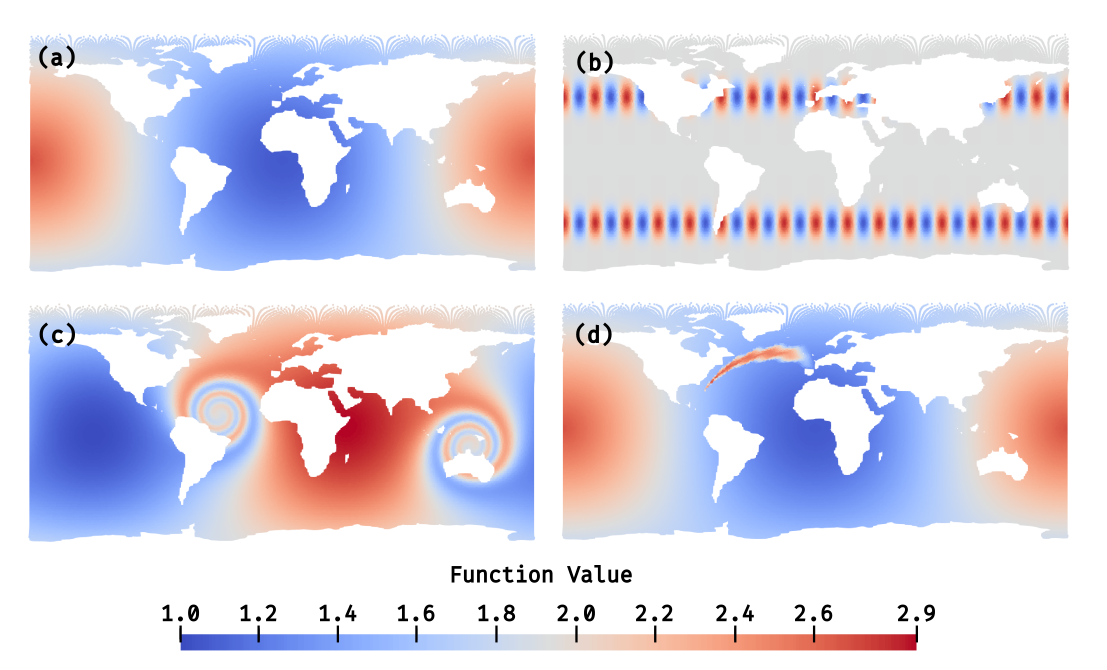}
    \caption{Lon/lat representation of nogt mesh with all test functions: a) sinusoid b) harmonic c) vortex d) gulfstream}
    \label{fig:eval-nogt}
\end{figure*}

\subsubsection{Data Mapping Methods}

The couplers used in the reference are SCRIP \citep{scrip}, YAC \citep{yac}, ESMF \citep{esmf}, and XIOS \citep{xios}.
The authors used a nearest-neighbor, a first-order, and a second-order method from each library--each one in a conservative and a non-conservative version using the ESM definition of conservation as described above. Table \ref{tab:naming-ref2} gives an overview of the used methods and the naming scheme we further use.

\begin{table}[H]
    \centering
    \begin{tabular}{c|c|c|c|c}
         our group name & \cite{paper}  & conservative        & non-conserv. & preCICE (non-conserv.) \\ \hline
         nearest-neighbor & nearest-neighbor  &      -              & distwgt-1 & nearest-neighbor (NN)       \\
         linear  & first order      & conserv-fracarea    & bilinear & nearest-projection (NP)       \\
         higher order & second order & cons2nd-fracarea    & bicubic & radial-basis function (RBF)\\
    \end{tabular}
    \vspace{0.2cm}
    \caption{Mapping methods used in original benchmark \citep{paper} and in our study}
    \label{tab:naming-ref2}
\end{table}

All four ESM couplers operate on the meshes defined by their nodes, whilst the data is assigned to cells, contrary to how data mapping is handled in preCICE.
Prior to any mapping, \cite{paper} map the SEA masks to the ATM meshes in a cell-local conservative manner. The resulting masks are then used during mapping, removing cells with mask values lower than $10^{-3}$. We follow a similar approach in Section \ref{sec:implementation} and give more details there.

\subsection{VTK Format}
\label{ssec:vtk}

Since the implementation presented in Section \ref{sec:implementation} relies on the VTK data format \citep{vtkBook}, a brief introduction is provided here. Data contained in VTK files may be unorganized in the form of data objects or have a certain underlying structure in the form of data sets. Our data set of choice is the unstructured grid (vtkUnstructuredGrid), which represents a collection of three-dimensional points and can have arbitrary topology. The later may be constructed as a combination of 0D (points), 1D (lines), 2D (triangles, quads,...) or 3D (tetrahedra,...) objects by specifying the respective indices of points. Each topological object can be referred to as a cell, to which one can also associate cell-data. Furthermore, one can attach point-data to points. As cells and points are indexed sequentially, associated data comes in form of data arrays, namely, vtkPointData or vtkCellData, respectively. Finally, data names allow distinguishing multiple data arrays.

\section{Implementation} \label{sec:implementation}

Since preCICE follows an application-agnostic coupling approach, the ESM benchmark meshes and data need to undergo specific pre- and post-processing. We face three particular challenges: First, as introduced above, preCICE handles meshes as pure point clouds and, thus, does not offer data mapping methods combining node locations with cell data. Second, for NP mapping, preCICE only supports triangles and quads as surface mesh elements, not more complex elements, such as hexagons. Last, also introduced above, we need to convert between Cartesian and geodesic coordinate systems.

Moreover, similarly to \cite{paper}, we need to transfer land-ocean masks from the SEA to the ATM meshes in a pre-processing step.
To increase the usability of the benchmark problem and the comparability of its results using different approaches, it could be meaningful to fix pre-computed masks to the ATM meshes in the definition of the benchmark.

In this section, we describe the six stages of our complete pipeline including pre-processing, the actual mapping, and post-processing. Figure \ref{fig:mesh-conversion-icos} visualizes the first three stages.
We end the section by describing mesh issues we observed in the original data and how we fixed them.

\begin{figure}[hbt!]
    \centering
    \includegraphics[width=0.7\linewidth]{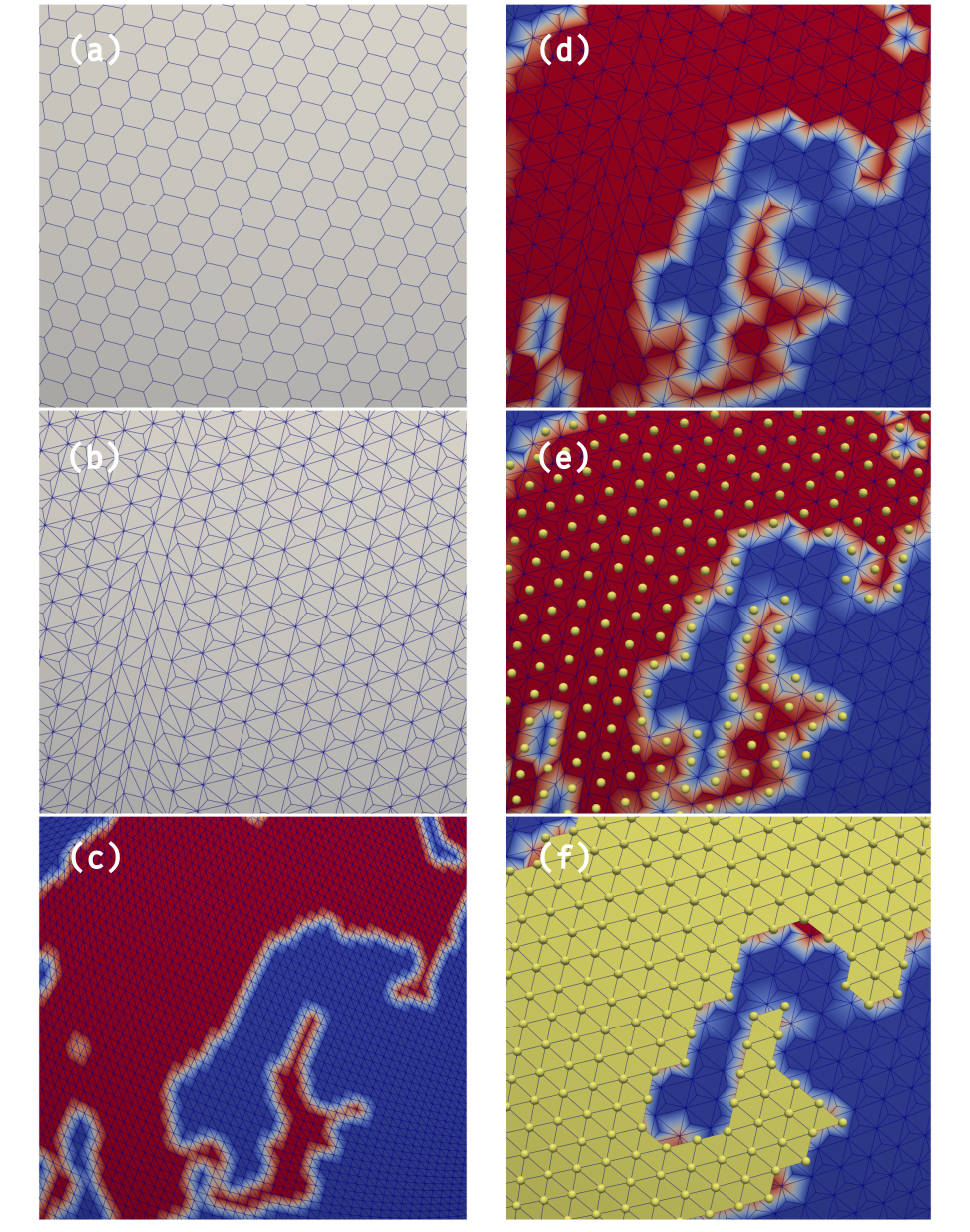}
    \caption{Mesh conversion process using a section of the ATM mesh icos and the SEA mesh nogt. a) icos, node-based b) icos, node-based \& triangulated c) nogt, node-based \& triangulated \& with mask d) icos, node-based \& triangulated \& mapped nogt mask e) icos, like d) with overlayed cell centers for each sea cell f) icos, like e) with added connectivity information for cell centers}
    \label{fig:mesh-conversion-icos}
\end{figure}

\subsection{SEA Masks}

To handle all pre-processing, we extract the meshes from the NC files, process them in Python, and write the results to VTK. Nodes and cell centers are converted from geodesic to Cartesian coordinates using the Proj library \citep{proj}. Duplicated points are merged below a precision of $10^{-8}$. The original SEA meshes provide the masks as cell data. To map them to the ATM meshes with preCICE, we first map them internally to the nodes by averaging the neighboring cell mask values. Each mesh now corresponds to one VTK file containing one \texttt{vtkUnstructuredGrids} mesh for the nodes (see Figure \ref{fig:mesh-conversion-icos}(a)). To the SEA meshes, we attach one \texttt{vtkPointData} array containing the mask.

Afterwards, we decompose all cells of all meshes using a Delaunay triangulation as preparation for the next stage. Figure \ref{fig:triangulation} sketches the necessary substeps. For each cell, we assume that its nodes $p_1, ..., p_N$ lay on a common plane.
We use the first three nodes to compute basis vectors $v_1$ and $v_2$ and the normal vector of the plane. Next, we sort all nodes counter-clockwise with respect to the normal vector and the center, to create triangles with normal vectors facing the same direction for visualization purposes.
We then project the nodes onto the plane defined by $v_1$ and $v_2$ to obtain two-dimensional coordinates for each nodes, which we finally use for the Delaunay triangulation (see Figure \ref{fig:mesh-conversion-icos}(b)).

\begin{figure}[hbt!]
    \centering
    \includegraphics[width=0.5\linewidth]{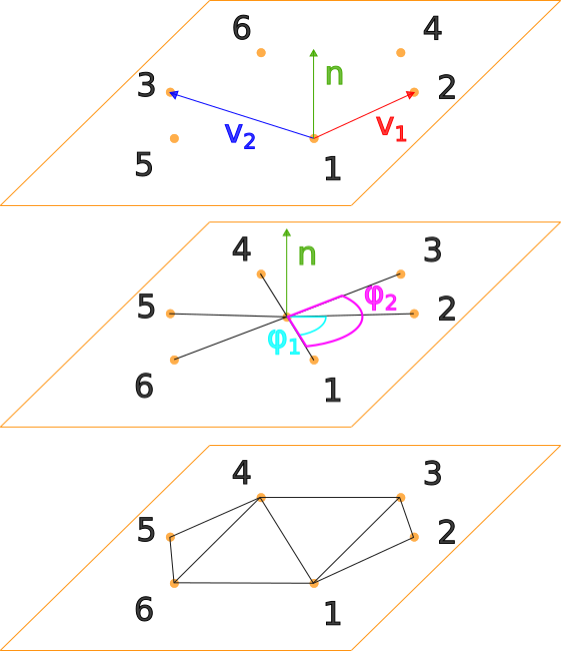}\\
    \caption{Substeps to add connectivity to meshes. From top to bottom: (i) calculation of plane basis vectors and normal, (ii) sorting of nodes around the cell center using normal and respective angles, (iii) result of Delaunay triangulation of nodes expressed in plane basis vectors}
    \label{fig:triangulation}
\end{figure}

\subsection{ATM Masks}

The original benchmark \citep{paper} uses a conservative mapping following their definition, as introduced above, to map the masks from SEA to ATM meshes. As preCICE does not yet offer such a mapping methods, we instead use a scaled-consistent NP mapping. We, thus, do not get the exact same masks as used by \cite{paper}, but only an approximation. The triangulation that we computed in the previous stage is required for the scaled-consistent NP mapping. In the end, we receive two masks for each ATM mesh, one for each of the two SEA meshes (see Figures \ref{fig:mesh-conversion-icos}(c) and \ref{fig:mesh-conversion-icos}(d)).

\subsection{Mesh Extraction}

We now have the mask data on the nodes of the ATM meshes (see Figure \ref{fig:mesh-conversion-icos}(d)). For the actual data mapping of the benchmark, the data sits on the cell centers, however. We, thus, need to decide whether we neglect a complete cell or not. To this end, we compute a cell mask value as the average of all surrounding nodes and neglect the cell (i.e. consider it land) if the mask values is smaller than $10^{-3}$ similarly to \cite{paper} (see Figure \ref{fig:mesh-conversion-icos}(e)). We write the cell centers and the resulting cell mask into further cell-based VTK files, which we ultimately use for the mapping tests.

As we want to evaluate the NP data mapping of preCICE, besides NN and RBF, we require connectivity information between cell centers for all meshes. We, thus, construct dual meshes by evaluating whether cells contain common nodes (see Figure \ref{fig:mesh-conversion-icos}(f)) and add the resulting connectivity information to the cell-based VTK files.
As a by-product of this search, we compute the area of the cells by summing up the areas of their triangles. These values are attached as a further \texttt{vtkPointData} array to the cell-based VTK files.

Finally, we attach a last \texttt{vtkPointData} array to each cell-based VTK file containing water fractions for all non-neglected cells. For SEA meshes, these fractions are all equal to one. For ATM meshes, we could use the mask values, but these would only give rough approximations for cells at the land-water interface. Instead, we compute the overlap of these cells with the cells of the associated SEA meshes. To this end, we first store all valid cell center locations of the SEA meshes in a SciPy KDTree space-tree data structure \citep{scipy}. For each ATM cell, we then query for neighboring SEA cells, compute the intersection areas, and add their ratios with respect to the ATM cell to the water fraction value.

\subsection{Function Evaluation}

To prepare the actual data mapping, we evaluate the test functions on the cell centers of all meshes and store them as another \texttt{vtkPointData} array in the cell-based VTK files.

\subsection{Mapping}

Similarly to the original benchmark, we compute data mappings between all combinations of SEA and ATM meshes in both directions. When mapping between a certain SEA and a certain ATM mesh, we only consider the ATM mesh that was created using this specific SEA mesh.
To simplify the comparison with \cite{paper}, we analogously only discuss and display a selection of 12 mesh combinations.

We test NN, NP, and RBF as data mapping variants of preCICE. 
For the RBF mapping, we use \textit{compact-polynomial-c6} as basis function \citep{preCICEv2}, requiring a support radius as additional parameter. To simplify our workflow and to test the robustness of the mapping, we use a single value of the parameter for all mesh combinations. In fact, we use $2\cdot 10^{6}$, which is approximately the size of nine cells of the icos mesh or 16\% of the Earth radius. For this choice, the basis functions cover enough cells for each source mesh while also not being too flat \citep{preCICEv2}. 
Alternatively, in future studies, we could make use of an automatic parameter optimization, which is currently developed for preCICE\footnote{\url{https://github.com/precice/precice/pull/2340}}.
Moreover, we set the relative overlap of the partition-of-unity clusters to 0.1 and use 350 vertices per cluster -- both slightly higher than their default values to increase robustness \citep{Schneider2025}.

\subsection{Metrics Evaluation}
\label{ssec:metrics}

As final step, we evaluate all mapping results based on the metrics introduced in Section \ref{sec:background}. As we need to evaluate the test functions on the target meshes, we need to convert all results to geodesic coordinates again. We add the misfit between computed and expected result as another \texttt{vtkPointData} array and store all derived metrics in a JSON file for further processing.

Similarly to \cite{paper}, we scale the stored surface area values with the water fraction when computing integral values. Note that this step adds some conservativness also to non-conservative mappings, but only in post-processing.

\subsection{Meshing Issues}

During initial tests, we observed two meshing issues in the benchmark data. These were presumably not caught by \cite{paper}, since either the ESM couplers offer different error handling measures -- such as \textit{-ignore\_degenerate} for ESMF to remove degenerated cells that, for example, collapsed into single lines or points -- or due to our different approach to map masks.

\begin{figure}[h!]
    \centering
    \includegraphics[width=0.6\linewidth]{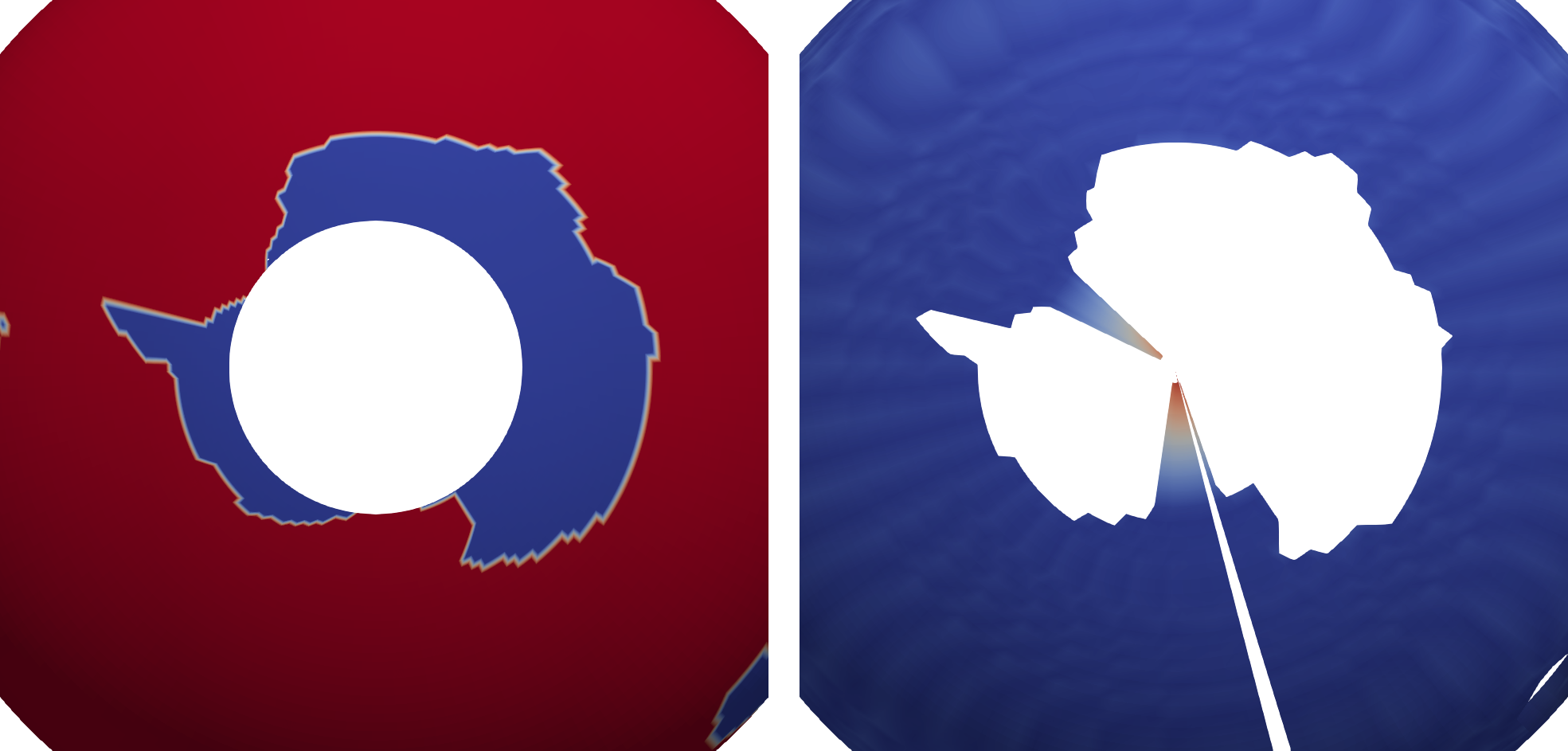}
    \caption{Malformed torc mesh around south pole. Torc mesh mask with missing cells in white (left) and resulting error on bggd mesh in cells not masked out correctly (right).}
    \label{fig:bad-torc-south}
\end{figure}

The first meshing issue are missing cells in the torc mesh for a circular region around the south pole. These are not unusual for a SEA mesh, but lead in our pre-processing to ATM cells not masked out correctly and a significant local increase of the misfit, as displayed in Figure \ref{fig:bad-torc-south}.
We solved this problem by hard-coding a solution in the pre-processing scripts: We added one node in the center of the circular region, apply Delaunay triangulation to add connectivity, and set the mask to land in the circle.

\begin{figure}[h!]
    \centering
    \includegraphics[width=0.6\linewidth]{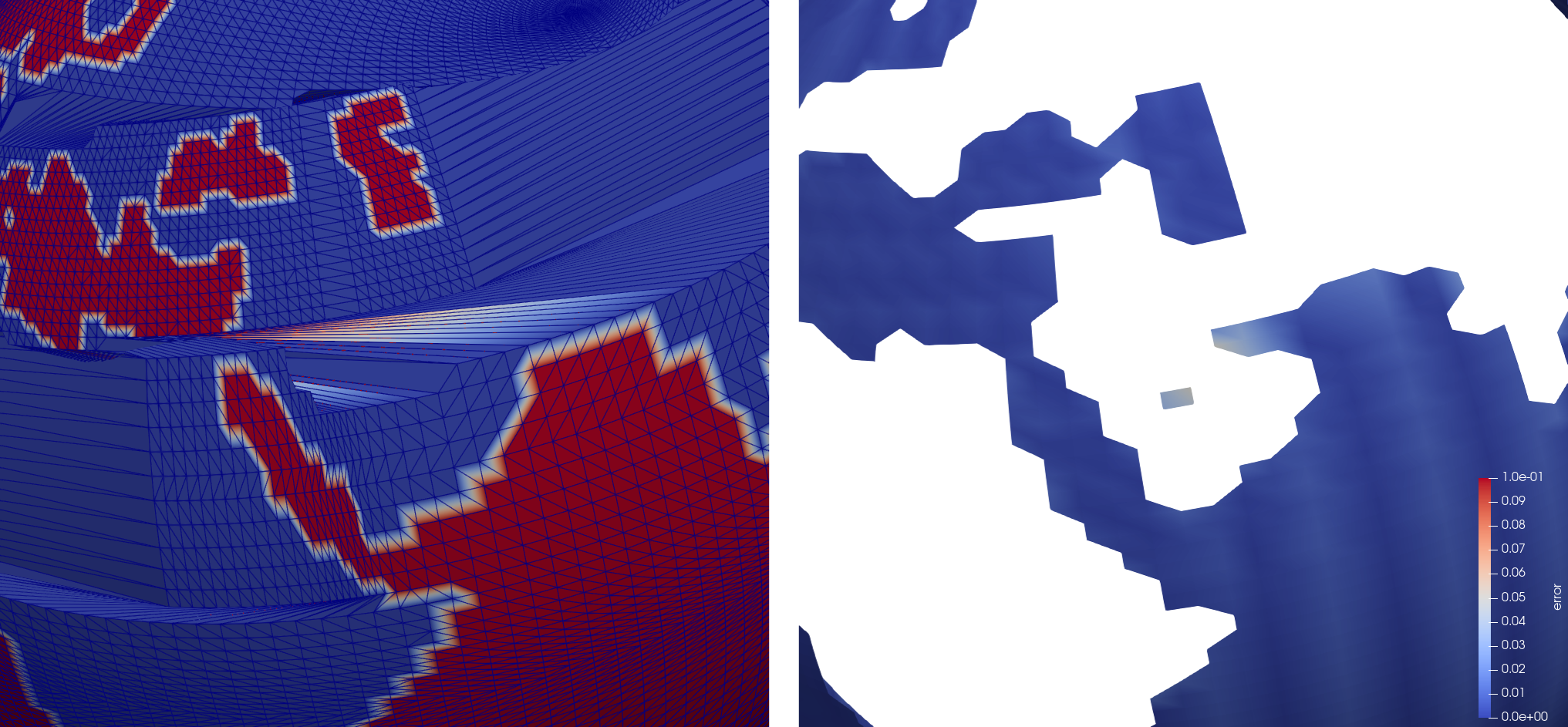}
    \caption{Malformed torc mesh in northern Africa. Torc mesh mask (left) and resulting error on bggd mesh (right).}
    \label{fig:bad-torc-africa}
\end{figure}

A similar issue arose in a region of the Mediterranean ocean. Here, malformed cells of the torc mesh connect regions in north Africa to Saudi-Arabia. This also resulted in cells not masked out correctly, see Figure \ref{fig:bad-torc-africa}. We removed the malformed cells in the pre-processing scripts.

\section{Results} \label{sec:results}

With the pipeline introduced in the previous section, we obtain all required metrics for the three preCICE mappings NN, NP, and RBF. We now compare these to the results of \cite{paper}, who tested the three couplers ESMF, SCRIP, and YAC. A comparison of these three among themselves is already covered by \cite{paper} and, thus, not repeated here.
In particular, we compare preCICE' NN mapping to the nearest-neighbor mappings of the reference, preCICE' NP mapping to the first-order mappings (our name: linear), and preCICE' RBF mapping to the second-order mappings (our name: higher order), see Table \ref{tab:naming-ref2}).

Since preCICE does not yet offer conservative mappings in the definition of \cite{paper}, we do not compare to the conservative results of the reference. 
In fact, the conservative mapping results of \cite{paper} do, in average, not show smaller source conservation errors than their non-conservative variants, only less variance. We presume that this is due to the scaling of the surface areas with the water fraction in post-processing as explained in Section \ref{ssec:metrics}.

\begin{figure*}[h!]
    \centering
    \includegraphics[width=0.95\linewidth]{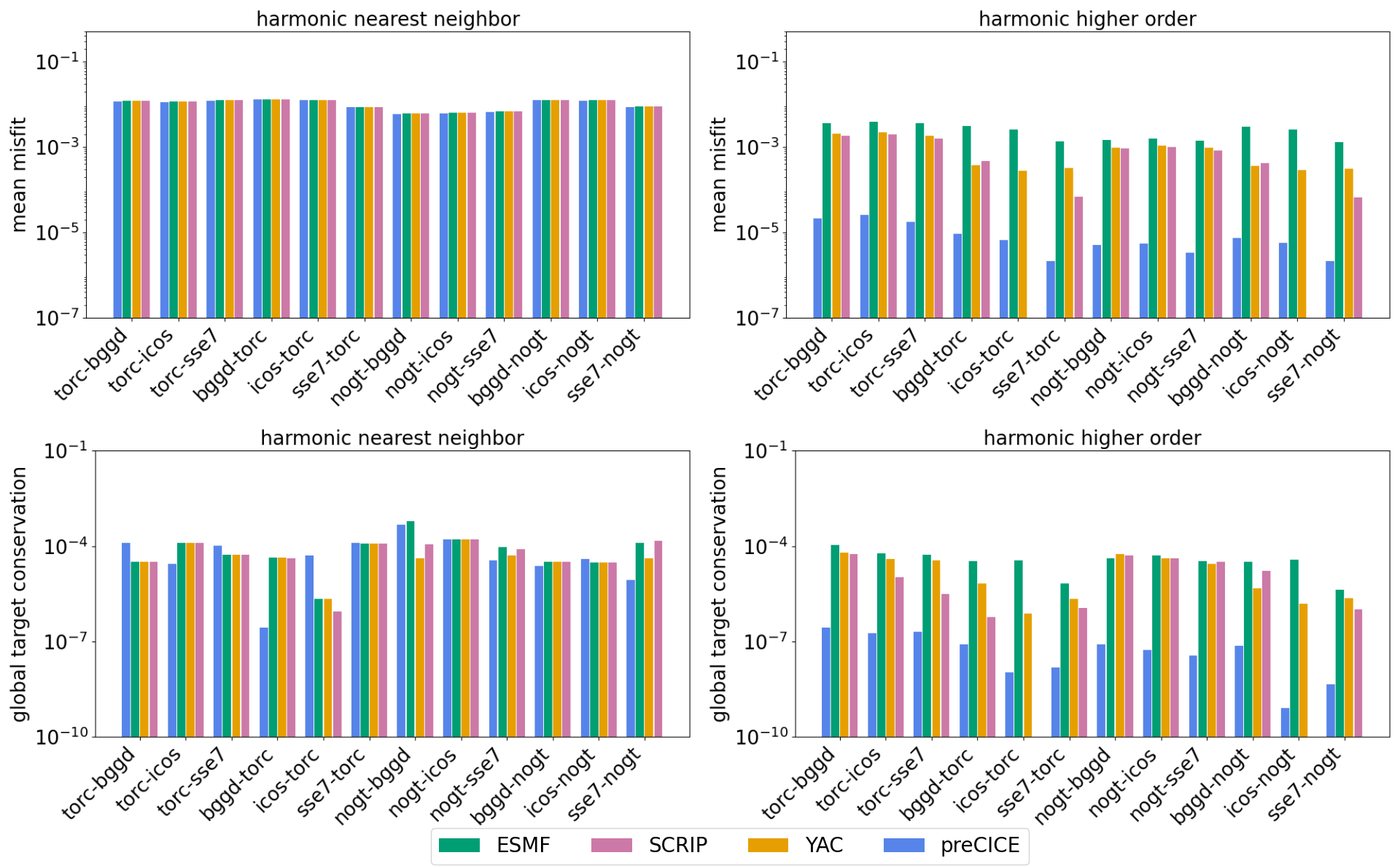}
    \caption{Overall representative behavior of consistent preCICE mapping compared to non-conservative reference.}
    \label{fig:general-cons}
\end{figure*}

In the shown figures and in the discussion, we restrict ourselves to the most representative and most interesting results. For full results and all details, we refer the reader to the associated data set~\citep{data}.
Figure \ref{fig:general-cons} shows the mean misfit and the global target conservation for the harmonic test function, for all considered mesh combinations, and for the nearest-neighbor and higher-order mappings. The misfit of preCICE' NN mapping shows very similar behavior to the reference. This also applies for all other test functions and also approximately for the linear mappings (not shown).
Overall, this validates our testing pipeline.

\begin{figure*}[h!]
    \centering
    \includegraphics[width=0.95\linewidth]{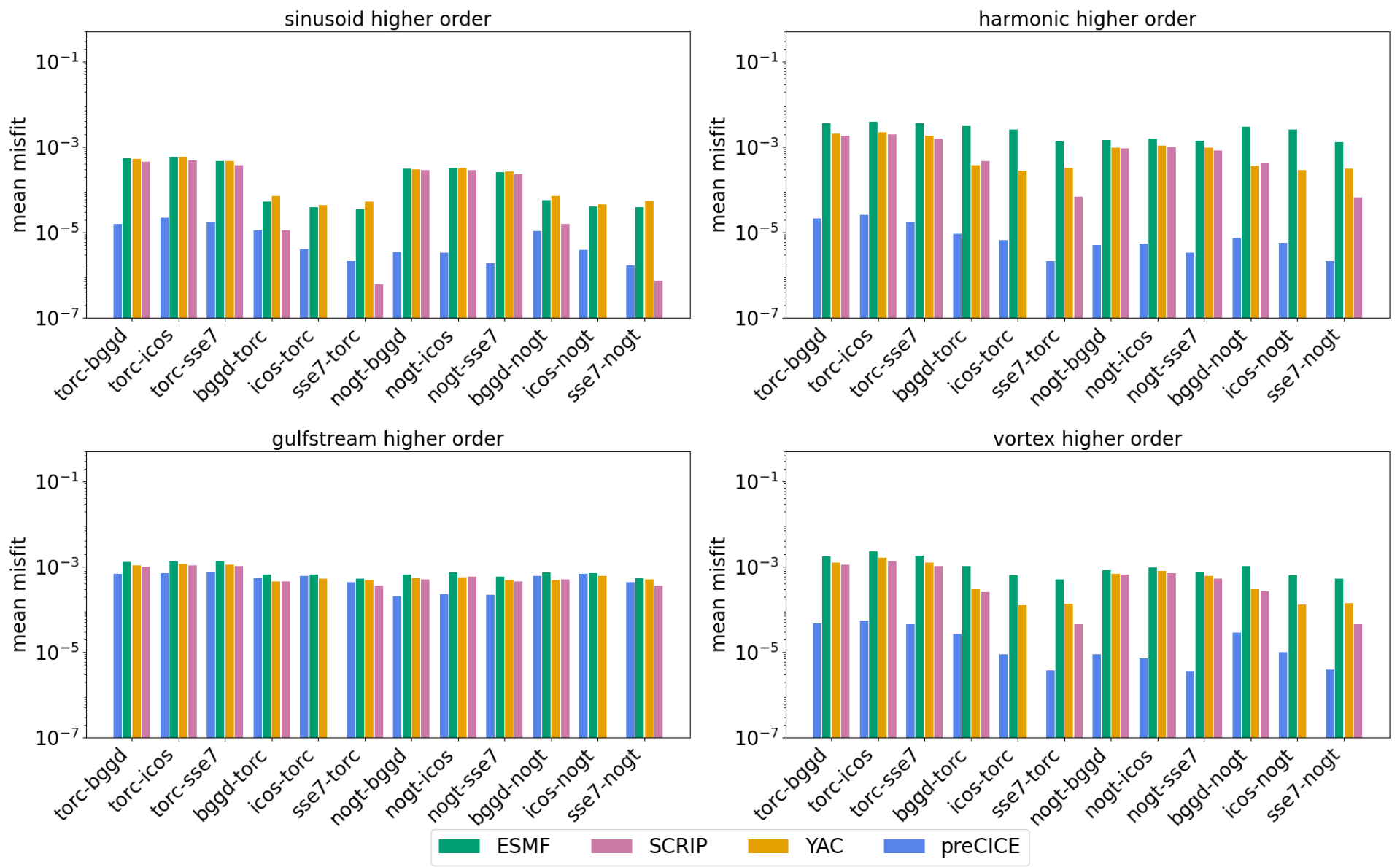}
    \caption{Overview of mean misfits across all test functions using higher order mapping methods. preCICE results are lower for all but the gulfstream case.}
    \label{fig:rbf-overview}
\end{figure*}

Comparing preCICE' RBF mapping to the higher order options of the reference, gives further insight, see also Figure \ref{fig:rbf-overview}. For all test functions except gulfstream, preCICE gives smaller misfits by two orders of magnitude. For gulfstream, we obtain comparable results to the reference. This can be explained by the high smoothness of all test functions other than gulfstream, for which RBF is known to perform well~\citep{Schneider2025}.

\begin{figure*}[h!]
    \centering
    \includegraphics[width=0.95\linewidth]{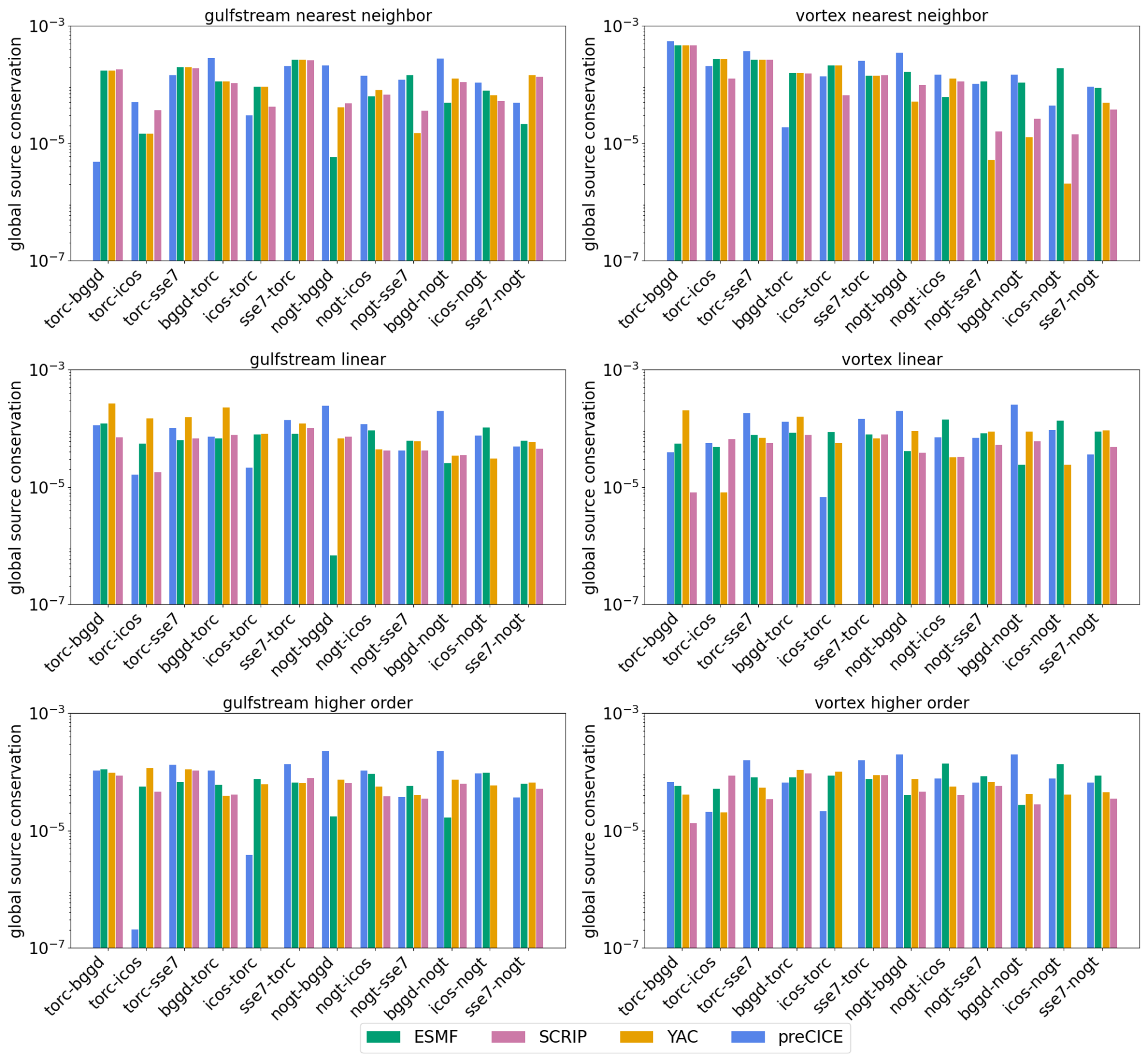}
    \caption{Global source conservation for all mapping methods and the gulfstream test function (left column) and the vortex test function (right column)}
    \label{fig:glob-cons-src-overview}
\end{figure*}

The global target conservation closely follows the misfit by definition and, thus, gives, as expected, similar results. More interesting is the source mesh conservation shown in Figure \ref{fig:glob-cons-src-overview}. For this metric, preCICE gives overall similar results than the reference. The slight differences can be observed for all test functions and all data mappings. Particularly, this is true for the nearest-neighbor mapping, which indicates that the differences are not due to the mapping method, but due to the computation of the metric during post-processing involving mesh connectivity.

The max-misfit metric behaves very similarly than the misfit metric. The results of preCICE follow the same trajectory as those of the reference, with the exception of the RBF mapping for smooth test functions. Also regarding the l-min and l-max metric, preCICE yields similar results as the reference without much deviation. For brevity, we do not show these results, but refer to the data set \citep{data}.

\begin{figure*}[h!]
    \centering
    \includegraphics[width=0.95\linewidth]{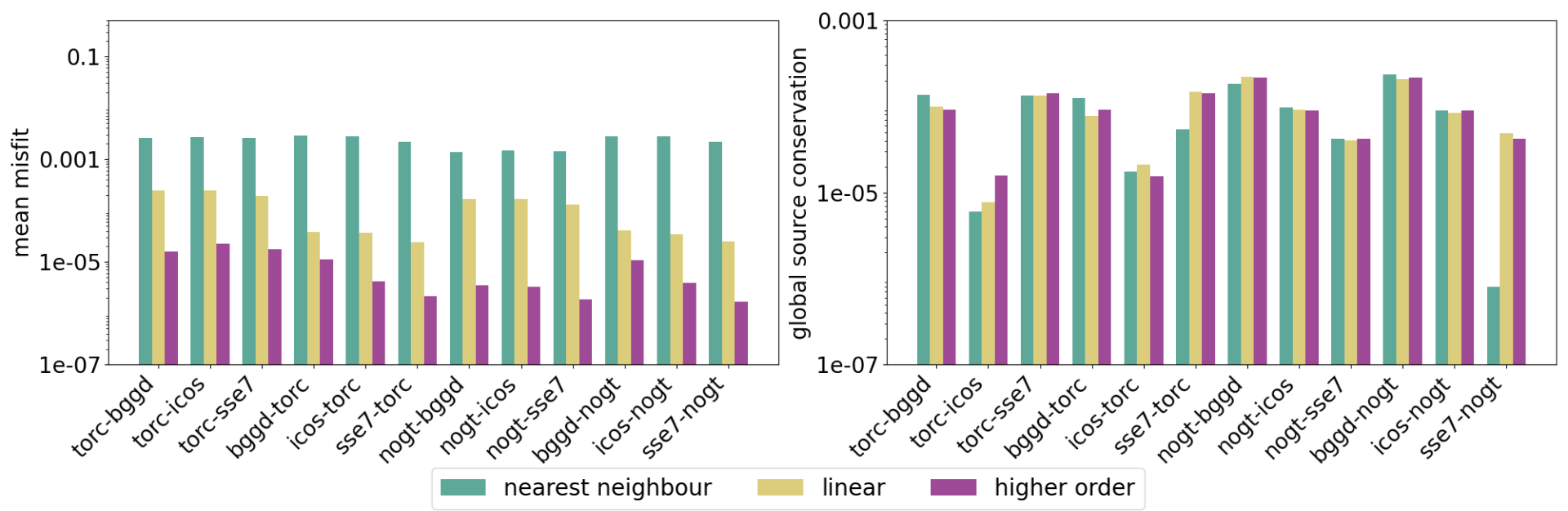}
    \caption{Overview of preCICE results using different mapping methods for the sinusoid test function.}
    \label{fig:metrics-ov-sin}
\end{figure*}

\begin{figure*}[h!]
    \centering
    \includegraphics[width=0.95\linewidth]{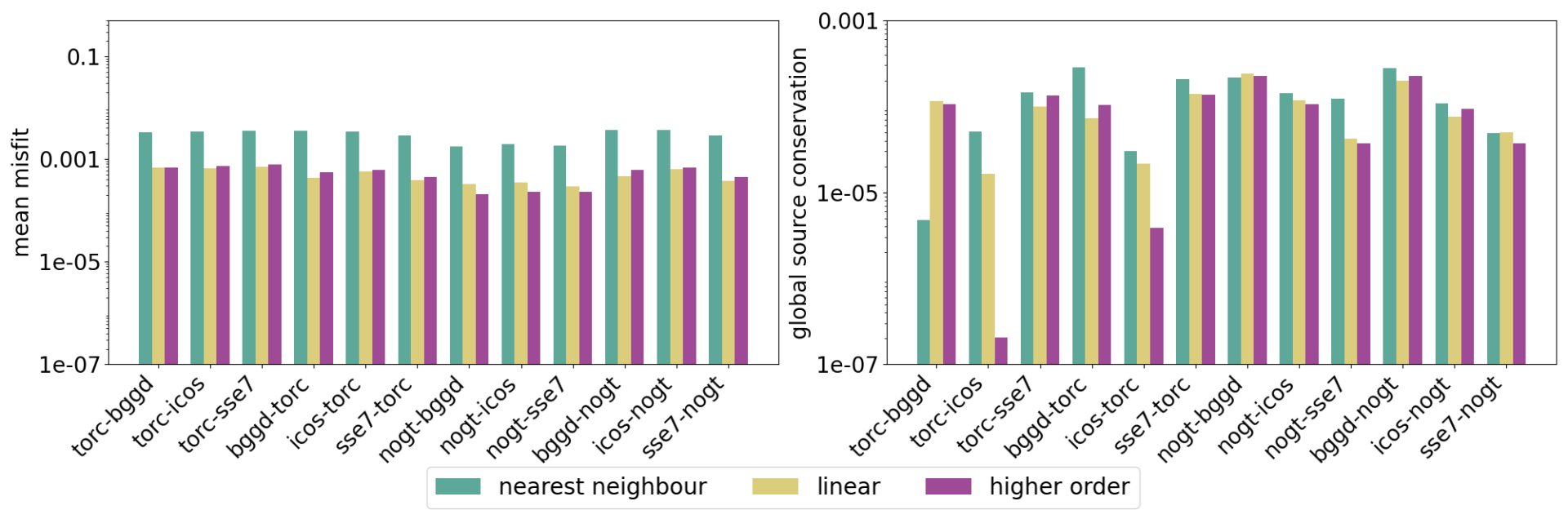}
    \caption{Overview of preCICE results using different mapping methods for the gulfstream test function.}
    \label{fig:metrics-ov-gul}
\end{figure*}

Finally, Figures \ref{fig:metrics-ov-sin} and \ref{fig:metrics-ov-gul} compare the three mapping methods of preCICE with each other for the smooth sinusoid function and the non-smooth gulfstream function, respectively.
For the sinusoid function, the misfit improves with the order of the data mapping method, while the global source conservation shows no dependence on the latter. For the gulfstream function, RBF gives no improvement over NP.

\conclusions
We evaluated the general-purpose coupling software preCICE in an Earth System Modeling (ESM) mapping benchmark \citep{paper}, which previously compared four ESM-specific couplers. Such a benchmark is highly valuable as it provides a common basis for comparison.

Overall, preCICE achieved comparable results, demonstrating that it can successfully handle ESM mapping problems. Its generality has led to a large and active user and developer community. This openness fosters cross-domain knowledge transfer, which can benefit ESM applications.
An example of such transfer is the RBF mapping method, widely used in fluid–structure interaction \citep{preCICEv2}. In our tests, it outperformed ESM-specific second-order methods by in average two orders of magnitude for smooth test functions while also showing robustness: The same setup worked across all considered ESM meshes.

However, this generality also comes at a cost: extensive pre- and postprocessing was required, partly due to the nature of ESM meshes and data and partly due to the benchmark's definition. Looking ahead, a detailed performance comparison to ESM-specialized couplers would be interesting, but lies beyond the scope of this paper as it would require a carefully refined benchmark definition.

\codedataavailability{}
We used: preCICE v3.3.0 (code and installation instructions: \url{https://precice.org/installation-overview.html}), ASTE v3.3.0 (code and installtion instruction: \url{https://precice.org/tooling-aste.html}), VTK v9.5.2 (installation from source is recommended, see ASTE documentation and \url{https://docs.vtk.org/en/latest/build_instructions/build.html#obtaining-the-sources}),
PROJ v9.7.0 (requires system-wide installation, alongside the python bindings, see:\\ \url{https://proj.org/en/stable/install.html#compilation-and-installation-from-source-code}). Frozen versions of the sources are archived on Zenodo under \url{https://doi.org/10.5281/zenodo.17897132} \citep{code}.

The current version of the scripting environment is available from the project repository \url{https://github.com/Snapex2409/ESM-NC-VTK-Tool} under the MIT licence. The exact version of the scripts used to produce the results used in this paper is archived on Zenodo under \url{https://doi.org/10.5281/zenodo.17368791} \citep{data}, as are input data and scripts to run the model and produce the plots for all the simulations presented in this paper \citep{data}.

\authorcontribution{}
AH: data curation, software, writing (original draft preparation),
BU: supervision, writing (review and editing)

\competinginterests{}
The authors declare that they have no conflict of interest.

\begin{acknowledgements}
We thank the authors of the reference paper S.~Valcke and G.~Jonville for providing further insights, technical support, and access to intermediate data as well as pre-release source code. The NetCDF software and dataformat from NSF Unidata was used to store the original datasets. We moreover want to thank D.~Abele, A.~Humbert, and T.~Kleiner for their help with the mesh and data formats and for their feedback on the manuscript.
We acknowledge the assistance of AI-tools (ChatGPT-5) for editorial suggestions.
The work of BU was funded by the Deutsche Forschungsgemeinschaft (DFG, German Research Foundation) under Germany's Excellence Strategy, EXC2075 -- 390740016. BU gratefully acknowledges the support by the Stuttgart Center for Simulation Science (SimTech).
\end{acknowledgements}

\bibliographystyle{copernicus}
\bibliography{ref.bib}

\end{nolinenumbers}
\end{document}